\documentclass[12pt]{book}
\usepackage{amssymb}
\usepackage{graphicx}
\usepackage{bm}
\newcommand{\bq}{\begin{equation}}
\newcommand{\eq}{\end{equation}}
\newcommand{\Ker}{\mathrm{Ker\,}}
\newcommand{\Rg}{\mathrm{Rg\,}}

\newcommand{\Thm}{\newtheorem}
\newcommand{\mathe}{{\mathrm e}}
\Thm{proposition}{Proposition}
\Thm{lemma}{Lemma}

\begin{document}
\chapter{Reduction of the chaotic transport of impurities in turbulent magnetized plasmas}
\label{sec:I}

\begin{center}
\Large
\textsc{C. Chandre$^1$, G. Ciraolo$^2$, M. Vittot$^1$}\\

\normalsize
\textit{ 
$^1$ Centre de Physique Th\'{e}orique
, CNRS - Aix-Marseille Universit\'es, Luminy, Case 907, F-13288 Marseille
cedex 9, France \\
$^2$ M2P2, IMT La Jet\'ee, Technop\^ole de Ch\^ateau Gombert, Marseille cedex 20, F-13451, France}
\end{center}

\section{Introduction}
\label{sec:IA}
The control of transport
in magnetically confined plasmas is of major importance in the long way to achieve
controlled thermonuclear fusion. Two major mechanisms have been proposed for such a
turbulent transport: transport governed by the fluctuations of the magnetic field and
transport governed by fluctuations of the electric field. There is presently a general
consensus to consider, at low plasma pressure, that the latter mechanism  agrees with
experimental evidence~\cite{scot03}. In the area of transport of trace impurities, i.e.
that are sufficiently diluted so as not to modify the electric field pattern, the {\em
$\bf E \times B$} drift motion of test particles should be the exact transport model. The possibility of reducing and even suppressing chaos
(combined with the empirically found states of improved confinement in tokamaks) suggest
to investigate the possibility to devise a strategy of control of chaotic transport
through some smart perturbations acting at the microscopic level of charged
particle motions.\\
Chaotic transport of particles advected by a turbulent electric
field with a strong magnetic field is associated with Hamiltonian dynamical systems under
the approximation of the guiding center motion due to ${\bf E \times B}$ drift velocity.
For an appropriate choice of turbulent electric field, it has been shown that the
resulting diffusive transport is found to agree with the experimental counterpart
\cite{pett88,pett89}. 

\begin{figure}
\centering
\includegraphics[width=0.8\textwidth]{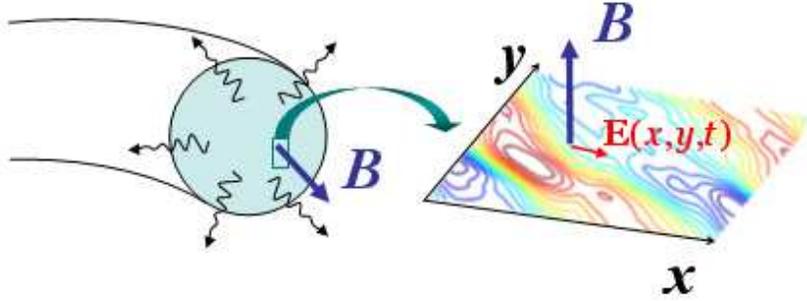}
\caption{The interaction between the electric turbulent field ${\bf E}$ and the toroidal magnetic field ${\bf B}$ produces the ${\bf E}\times{\bf B}$ motion which causes a drift of particles from the center of the plasma toward the edge. The guiding center motion of charged test particles on a transversal plane to the toroidal direction can be described by a Hamiltonian formalism with the spatial coordinates $(x,y)$ which represents canonical conjugate variables and with the electrostatic potential $V$, ${\bf E}=-{\bm\nabla}V$, which plays the role of the Hamiltonian, as described by Eq.~(\ref{guidcent}).}
\label{FigIIIB_0}
\end{figure}
 
Here we address the turbulent transport of particles governed by
the interaction of the electric turbulent potential ${\bf E}$
generated by the plasma itself and the strong confining magnetic
field ${\bf B}$ produced by external coils. The resulting ${\bf E \times B}$ drift motion is perpendicular to the
confining magnetic field and causes losses of particles and energy
from the inner region to the border of the device, with a
consequent decrease of plasma temperature (see Fig.~\ref{FigIIIB_0}). The equations of this
drift motion for charged test particle, in the guiding center
approximation, have a Hamiltonian structure and are given by
\begin{equation}
\label{guidcent}
{\dot{\bf x}}={\mathrm{d}\over \mathrm{d}t}{x\choose y}={c\over B^2}{\bf E}({\bf x},t)\times
 {\bf B}=
{c\over B}{-\partial_y V (x,y,t)\choose \partial_x V (x,y,t)},
\end{equation}
where ${\bf x}=(x,y)$ represents the spatial coordinates of the transversal
section to the confining toroidal magnetic field, $B$ the norm of the magnetic field ${\bf B}$, 
$c$ the velocity
of light and $V$ is the electric turbulent potential, that is
${\bf E}=-{\bm\nabla} V$. We notice that the resulting dynamics is of Hamiltonian nature with a pair of canonically conjugate variables $(x,y)$ which consists of the position of the guiding center. Since the potential which plays the role of the Hamiltonian is time-dependent, it is expected that the dynamics is chaotic (with one and a half degrees of freedom).\\

The sensitivity of chaotic systems to small perturbations
triggered a strong interdisciplinary effort to control chaos~\cite{Blich83,Bkada98,Bchen98,gaut03,ditt90,petr93,schi94,brai95}.
After the seminal work on optimal control by
Pontryagin~\cite{Bpont61}, new and efficient methods were proposed
for controlling chaotic systems by nudging targeted trajectories~\cite{ott90,lima90,fron91,shin93,ott95,lima98}. However, for many body experiments like
the magnetic confinement of a plasma or the control of
turbulent flows, such methods are hopeless due to the high
number of trajectories to deal with simultaneously.

Here the control is performed with the addition of a control term $f$ to the electric potential. It consists as a small and apt
modification of the electric potential with relevant effects on
the reduction of chaotic particle diffusion. More generally, controlling Hamiltonian systems here means to achieve the goal
of suppressing chaos when it is harmful as well as to enhance chaos
when it is useful, by driving
the dynamical evolution of a system toward a
``target behavior'' by means of small and apt perturbations. For magnetic confinement devices for controlled
thermonuclear fusion, the two possibilities of increasing chaos and suppressing
it are of interest. In fact, in the core of a magnetic confinement
device the aim is to have a stable and regular dynamics in order to
induce as many fusion reactions as possible, that is a high rate of energy
production.
At the same time, in order to spread heat over a large area, which is
fundamental both for a good
recycling of energy and for the preservation of plasma facing components,
at the edge of the confinement device the problem of draining highly
energetic particles requires the capacity of increasing the degree of
chaoticity of the dynamics.

For reducing chaos, KAM theory gives a path to integrability and the possibility of
controlling Hamiltonian systems by modifying
it with an apt and small perturbation which preserves the Hamiltonian
structure. In fact, if on one hand it is evident that perturbing an
integrable system with a
generic perturbation gives rise to a loss of stability and to
the break-up of KAM tori, on the other hand the structure of the Hamiltonian
system changes on the set of KAM tori with a continuity
$C^{\infty}$~\cite{posc82}
with respect to the amplitude of the perturbation.
The idea is that a well selected perturbation instead of producing chaos
makes the system integrable or more regular, recovering structures
as the KAM tori present in the integrable case.

Starting from these ideas we have addressed
the problem of control in Hamiltonian systems. For a wide class of chaotic
Hamiltonians, expressed as perturbations of integrable Hamiltonians, that is
$H_0+\varepsilon V$, the aim is to design a control term $f$ such that
the dynamics of the controlled Hamiltonian $H_0+\varepsilon V+f$
has more regular trajectories (e.g.~on invariant tori) or less chaotic
diffusion than the uncontrolled one. Obviously $f=-\varepsilon V$ is a solution
since the
resulting Hamiltonian is integrable. However, it is a useless solution
since the control is of the same magnitude of the perturbation while,
for energetic purposes, the desired control term should be small
with respect to the perturbation $\varepsilon V$. For example the control term should be of
order $\varepsilon^2$ or higher. In many physical situations, the control is only
conceivable or interesting in some specific regions of phase space where the
Hamiltonian can be known and/or an implementation is possible.
Moreover, it is desirable to control the transport
properties without significantly altering the original setup of
the system under investigation nor its overall chaotic structure.
The possibility of recreating specific regular structures in phase space such as
barriers to diffusion, can be used to bound the motion of particles without
changing the phase space on both sides of the barrier.

In addition, the construction of the control term has to be robust.
In fact, one expects from a KAM theorem also a given robustness
with respect to modifications of the well selected and regularizing
perturbation. This fact still comes from the previous results that we
mentioned above about the transition from integrability
to chaos that Hamiltonian systems generically show.
In other words, if an apt modification of a chaotic Hamiltonian gives rise to
a regularization of the dynamics, with for example the creation of KAM
tori, one expects that also some approximations of this apt modification
are still able to perform a relevant regularization of the system.
This is a fundamental requirement for any kind of control scheme
in order to guarantee the stability of the results and
in view of possible experimental implementations.

\section{Control term for Hamiltonian flows}
\label{sec:IC}

In what follows, we explain a method to compute control terms of order $\varepsilon^2$ where $\varepsilon$ is the amplitude of the perturbation, which aims at restoring the stable structures which were present in the case where $\varepsilon=0$. We notice that such stable structures are present for Hamiltonian systems $H_0+\varepsilon^{\prime} V$
where $\varepsilon^{\prime} < \varepsilon_c < \varepsilon$ (where $\varepsilon_c$ is the threshold of break-up of the selected invariant torus), even though they are deformed (by an order $\varepsilon^{\prime}$).

For a function $H$, let $\{H\}$ be the linear operator such that
$$
\{H\}H^{\prime}=\{H,H^{\prime}\},
$$
for any function $H^{\prime}$, where $\{\cdot~,\cdot\}$ is the Poisson bracket. 
The time-evolution of a function $V$ following the flow of $H$ 
is given by
$$
\frac{dV}{dt}=\{ H\} V,
$$  
which is formally solved as
$$
V(t)=\mathe^{t\{H\}}V(0),
$$
if $H$ is time independent.
Let us now consider a given Hamiltonian $H_0$.
The operator
\(\{{H_0}\} \) is not invertible since a derivation has always 
a non-trivial kernel (which is the set of constants
of motion).
Hence we consider a pseudo-inverse of \( \{{H_0}\} \).
We define a linear operator $\Gamma$ such that
\bq
\{{H_0}\}^{2}\ \Gamma = \{{H_0}\},
\label{gamma}
\eq
i.e.
$$
\forall V, \qquad \{H_0,\{H_0,\Gamma V\}\}=\{H_0,V\}.
$$
The operator $\Gamma$ is not unique (see the remark at the end of this section). 
\\ \indent We define the {\em non-resonant} operator $\mathcal N$ and the 
{\em resonant} operator $\mathcal R$ as
\begin{eqnarray*}
&& {\mathcal N} = \{H_0\}\Gamma,\\
&& {\mathcal R} = 1-{\mathcal N},
\end{eqnarray*}
where the operator $1$ is the identity. 
We notice that the range \( \Rg \mathcal R \) of the operator \( \mathcal R \) is
included in \( \Ker\{{H_0}\} \).
A consequence
is that ${\mathcal R} V$ is constant under the flow of $H_0$. 
\\ \indent Let us now assume that $H_0$ is integrable 
with action-angle variables 
$({\bf A},\bm{\varphi})\in {\mathbb B}\times {\mathbb T}^L $ where ${\mathbb B}$ 
is an open set of ${\mathbb R}^L$ and ${\mathbb T}^L$ is the $L$-dimensional torus. Thus 
$H_0=H_0({\bf A})$ and the Poisson bracket $\{H,H^{\prime}\}$ between two elements $H$ and $H'$ of ${\mathcal A}$ is 
$$
\{H,H^{\prime}\}=\frac{\partial H}{\partial{\bf A}}\cdot
\frac{\partial H^{\prime}}{\partial{\bm{\varphi}}}-
\frac{\partial H}{\partial{\bm{\varphi}}}\cdot
\frac{\partial H^{\prime}}{\partial{\bf A}}.
$$
The operator $\{H_0\}$ acts on $V$ expanded as follows
$$
V=\sum_{{\bf k}\in {\mathbb Z}^L}V_{\bf k}({\bf A})\mathe^{i{\bf k}\cdot{\bm\varphi}},
$$
as
$$
\{H_0\}V({\bf A},\bm{\varphi})=\sum_{\bf k}i{\bm \omega}({\bf A})\cdot{\bf k}~V_{\bf k}({\bf A})\mathe^{i{\bf k}\cdot\bm\varphi},
$$
where 
$$
{\bm \omega}({\bf A})=\frac{\partial H_0}{\partial{\bf A}}. 
$$
A possible choice of $\Gamma$ is
\begin{equation}
\label{eqn:GV}
\Gamma V({\bf A},\bm{\varphi})=
\sum_{{\bf k}\in{{\mathbb Z}^L}\atop{\omega({\bf A})\cdot{\bf k}\neq0}}
\frac{V_{\bf k}({\bf A})}
{i{\bm \omega}({\bf A}) \cdot{\bf k}}~~\mathe^{i{\bf k}\cdot{\bm\varphi}}.
\end{equation}
We notice that this choice of $\Gamma$ commutes with $\{H_0\}$. For a given $V\in{\mathcal A}$, ${\mathcal R} V$ is the resonant 
part of $V$ and ${\mathcal N} V$ is the non-resonant part:
\begin{eqnarray}
&&{\mathcal R}V=\sum_{{\bm\omega}({\bf A})\cdot {\bf k}=0}
V_{\bf k}({\bf A})\mathe^{i{\bf k}\cdot{\bm\varphi}},\label{eqn:RV}\\ 
&&{\mathcal N}V=\sum_{{\bm\omega}({\bf A})\cdot {\bf k}\not= 0}
V_{\bf k}({\bf A})\mathe^{i{\bf k}\cdot{\bm\varphi}}.\label{eqn:NV}
\end{eqnarray}
From these operators defined from the integrable part $H_0$, we construct a control term for the perturbed Hamiltonian $H_0+V$ where $V\in {\mathcal A}$, i.e.
 $f$ is constructed such that $H_0+V+f$
is canonically conjugate to $H_0+\mathcal R V$.
We have the following equation
\bq
\mathe^{\{\Gamma V\}}(H_0+V+f)=H_0+{\mathcal R} V,
\label{prop1}
\eq
where
\bq
f=\mathe^{-\{\Gamma V\}}{\mathcal R}V+\frac{1-\mathe^{-\{\Gamma
V\}}}{\{\Gamma V\}} {\mathcal N} V -V.
\eq
We notice that the operator $(1-\mathe^{-\{\Gamma V\}})/\{\Gamma V\}$
is defined by the expansion
$$
\frac{1-\mathe^{-\{\Gamma V\}}}{\{\Gamma V\}}=
\sum_{n=0}^{\infty}\frac{(-1)^n}{(n+1)!}\{\Gamma V\}^n.
$$
The control term can be expanded in power series as
\bq
f=\sum_{n=1}^{\infty}\frac{(-1)^n}{(n+1)!}\{\Gamma V\}^n
(n{\mathcal R}+1)V.
\label{expansion_f}
\eq
\\
We notice that
if $V$ is of order $\epsilon$, $f$ is of order $\epsilon^2$.
Therefore the addition of a well chosen 
control term $f$ makes the Hamiltonian canonically
conjugate to $H_0+{\mathcal R} V$. 

\indent If $H_0$ is non-resonant then with the addition of a
control term $f$, the Hamiltonian $H_0+V+f$ is 
canonically conjugate to the integrable Hamiltonian $H_0+{\mathcal R} V$
since ${\mathcal R} V$ is only a function of the 
actions [see Eq.~(\ref{eqn:RV})]. If $H_0$ is resonant and ${\mathcal R} V=0$, the controlled
Hamiltonian $H=H_0+V+f$ is conjugate to $H_0$. In the case ${\mathcal R} V=0$, the series~(\ref{expansion_f})
which gives the expansion of the control term $f$,
can be written as
\bq
f=\sum_{s=2}^{\infty}f_s,
\label{exp_f_rv_0}
\eq
where $f_s$ is of order $\varepsilon^s$, where $\varepsilon$ is the size of the perturbation $V$, and is given by the
recursion formula
\bq
f_s=-\frac{1}{s}\{\Gamma V,f_{s-1}\},
\label{recursion}
\eq
where $f_1=V$.

{\em Remark~: Non-unicity of $\Gamma$}-- The operator $\Gamma$ is not unique. Any other choice
$\Gamma^{\prime}$ satisfies that the range 
$\rm{Rg}(\Gamma^{\prime}-\Gamma)$ is included into the kernel
$\rm{Ker}(\{H_0\}^2)$. It is straightforward to see that adding a constant to $\Gamma$ does not change the expression of the control term. However, different control terms are obtained by adding a more complex linear operator to $\Gamma$. For instance, we choose
$$
\Gamma^\prime=\Gamma +{\mathcal R}{\mathcal B},
$$
where ${\mathcal B}$ is any linear operator. The operator $\Gamma^\prime$ satisfies $\{H_0\}^2\Gamma^\prime=\{H_0\}$ since $\{H_0\}{\mathcal R}=0$. The control term we obtain is given by Eq.~(\ref{expansion_f}) where $\Gamma$ is replaced by $\Gamma^\prime$. We notice that the new resonant operator defined as ${\mathcal R}^\prime=1-\{H_0\}\Gamma^\prime$ is equal to ${\mathcal R}$ since $\{H_0\}{\mathcal R}=0$. The new control term is in general different from $f$. For instance, its leading order is (still of order $\varepsilon^2$)
$$
f_2^\prime=f_2-\frac{1}{2}\{{\mathcal R}{\mathcal B} V, ({\mathcal R}+1)V\}.
$$
This freedom in choosing the operator $\Gamma$ can be used to simplify the control term (for withdrawing some Fourier coefficients) or more generally, to satisfy some additional constraints. 

{\em Remark~: Higher-order control terms}-- It is possible to construct higher order control terms, i.e.\ control terms of order $\varepsilon^n$ with $n>2$, and there are many ways to do so. In this paragraph, we add $\varepsilon$ in the control term for bookkeeping purposes. For instance, we notice that $H_0^{(c)}=H+\varepsilon V+\varepsilon^2 f$ is integrable. The perturbed Hamiltonian can be written as
$$
H=H_0^{(c)}-\varepsilon^2 f.
$$  
For simplicity, we assume a non-resonant condition on $V$, i.e.\ ${\mathcal R}V=0$. We define a new operator $\tilde\Gamma$ as
$$
\tilde{\Gamma}=\mathe^{-\varepsilon\{\Gamma V\}}\Gamma\mathe^{\varepsilon\{\Gamma V\}},
$$
which is $\varepsilon$-close to $\Gamma$. It is straightforward to check that $\tilde{\Gamma}$ satifies
$$
\{H_0^{(c)}\}^2\tilde{\Gamma}=\{H_0^{(c)}\},
$$
since we have $\{H_0^{(c)}\}=\mathe^{-\varepsilon\{\Gamma V\}}\{H_0\}\mathe^{\varepsilon\{\Gamma V\}}$ from Eq.~(\ref{prop1}). By applying again Eq.~(\ref{prop1}) and replacing $V$ by $-\varepsilon^2 f$ and $H_0$ by $H_0^{(c)}$ (and consequently $\varepsilon$ by $-\varepsilon^2$), we have the existence of a control term $\varepsilon^4 g$ which satisfies~:
$$
H_0+\varepsilon V+\varepsilon^4 g=\mathe^{\varepsilon^2\{\tilde{\Gamma} f\}}\left( H_0^{(c)}-\varepsilon^2\tilde{{\mathcal R}}f \right),
$$
where $\tilde{\mathcal R}$ is the resonant operator associated with $\tilde{\Gamma}$ defined as $\tilde{\mathcal R}=1-\{H_0^{(c)}\}\tilde{\Gamma}$ (and is $\varepsilon$-close to ${\mathcal R}$). It follows from the commutation of $\{\tilde{\mathcal R}V\}$ and $\{H_0^{(c)}\}$ that the controlled Hamiltonian $H_0+\varepsilon V+\varepsilon^4 g$ is integrable. An expansion in $\varepsilon$ of the control term $g$ gives its leading order
$$
g_4=-\frac{\varepsilon^4}{8}\{\Gamma\{\Gamma V,V\},\{\Gamma V,V\}\}.
$$  
However, we notice that $\Gamma\{\Gamma V,V\}$ introduces additional small denominators. It is not obvious that, given a value of $\varepsilon$, the control term $\varepsilon^4 g$ is smaller than $\varepsilon^2 f$ (using a standard norm of functions).

\section{Reduction of chaotic transport for ${\bf E}\times {\bf B}$ drift in magnetized plasmas}
\label{sec:IIIB} 
\subsection{A paradigmatic model of electric potential}
\label{sec:IIIB1}

We consider the following model of electrostatic potential~\cite{cira04b}
\begin{equation}
V ({\bf x},t)=\sum_{{\bf k}\in{\mathbb Z^2}}{V_{\bf k}}\sin \left
[ \frac{2\pi}{L}{\bf k}\cdot {\bf x} +\varphi_{\bf k}-\omega ({\bf
k})t\right ],
\end{equation}
where $\varphi_{\bf k}$ are random phases (uniformly distributed)
and $V_{\bf k}$ decrease as a given function of $\Vert {\bf
k}\Vert$, in agreement with experimental data \cite{woot90}, and $L$ is the typical size of the elementary cell as represented in Fig.~\ref{FigIIIB_0}. In
principle one should use for $\omega ({\bf k})$ the dispersion
relation for electrostatic drift waves (which are thought to be
responsible for the observed turbulence) with a frequency
broadening for each ${\bf k}$ in order to model the experimentally
observed spectrum $S({\bf k},\omega)$.

\begin{figure}
\centering
\includegraphics[width=0.49\textwidth]{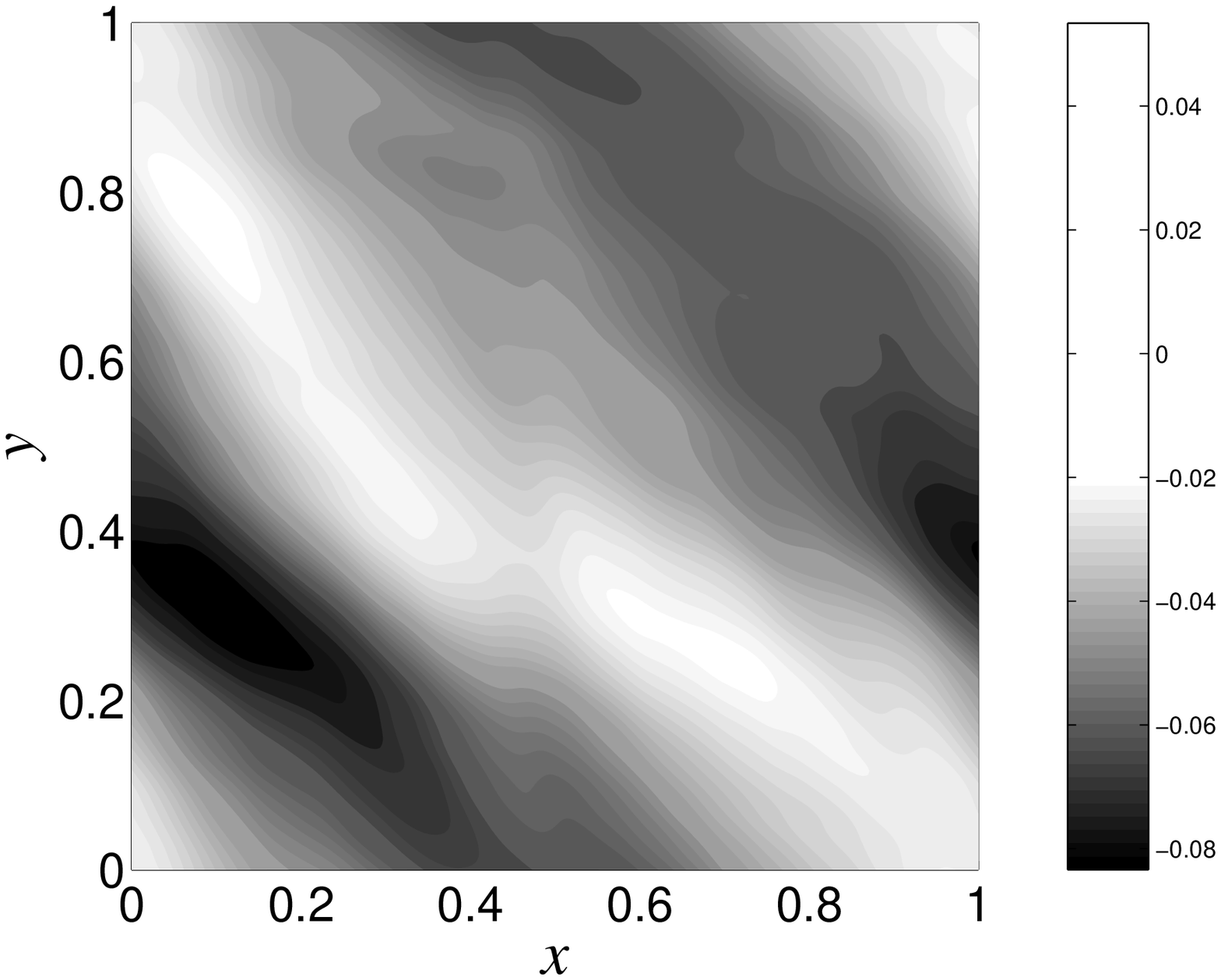}
\includegraphics[width=0.49\textwidth]{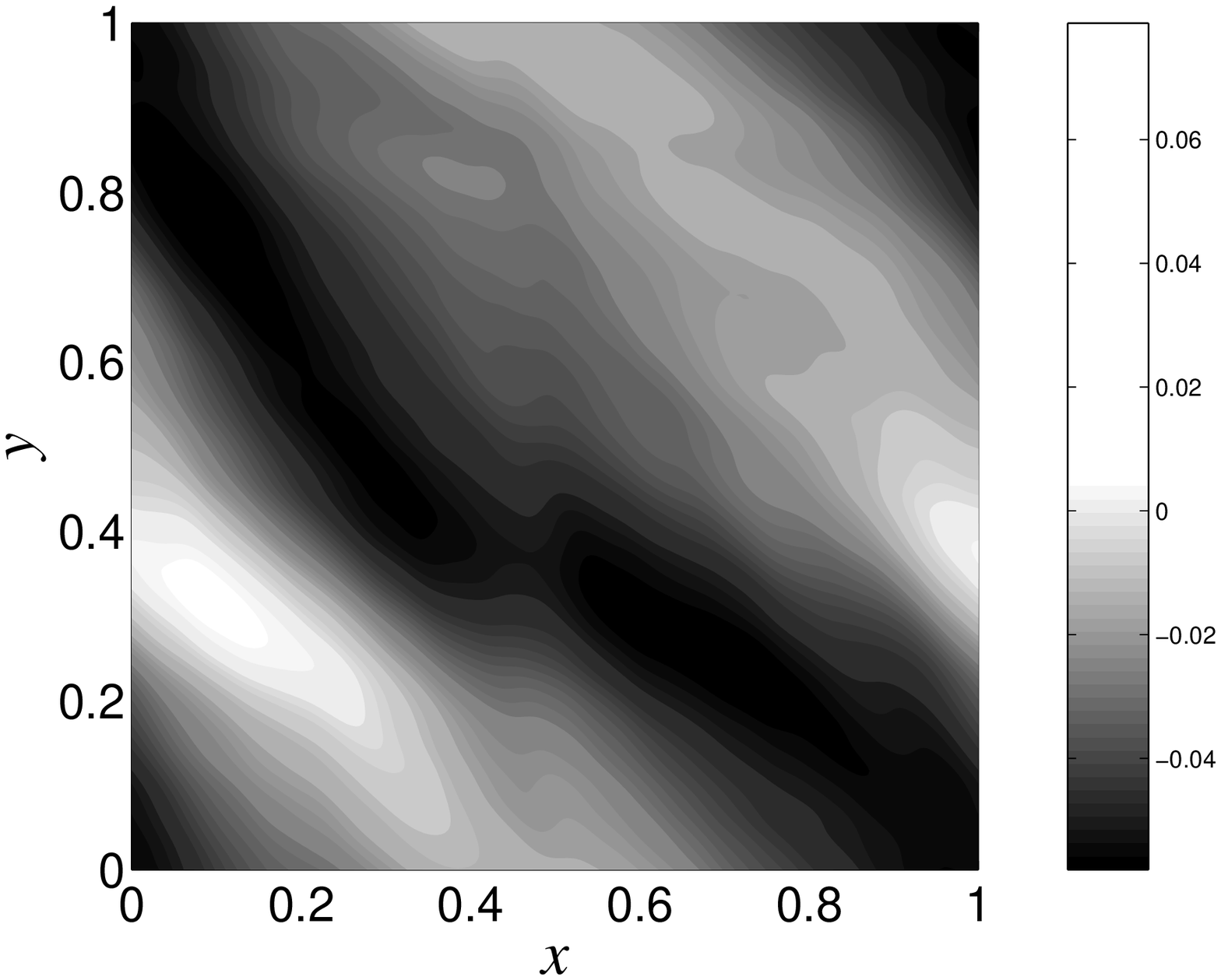}
\caption{Contour plots of $V(x,y,t)$  given by
Eq.~(\ref{potential}) for $a=1$, $N=25$, $t=0$ (left panel) and $t=1/2$ (right panel), .}
\label{FigIIIB_1}
\end{figure}

Here we consider a quasiperiodic approximation of the turbulent
electric potential with a finite number $K$ of frequencies. We
assume that $\omega_k\not= 0$ (otherwise, see remark at the end of Sec.~\ref{sec:IIIB2}). The phases
$\varphi _{\bf k}$ are chosen at random in order to mimic  a
turbulent field with the reasonable hope that the properties of
the realization thus obtained are not significantly different from
their average. In addition we take for $\Vert{V_{\bf k}}\Vert$ a
power law in $\Vert{\bf k}\Vert$ to reproduce the spatial spectral
characteristics of the experimental $S({\bf k})$, see
Ref.~\cite{woot90}. Thus we consider the following explicit form
of the electric potential:
\begin{equation}
\label{eqn:Vqp} V(x,y,t)=\sum_{k=1}^K\sum_{m,n=1}^N
\frac{a_k}{2\pi(n^2+m^2)^{3/2}} \sin [ 2\pi(nx+my)+\varphi_{kmn}-2\pi\omega_k t
],
\end{equation}
where $\varphi_{kmn}$ are random phases.
\\
For example, Fig.~\ref{FigIIIB_1} shows contour plots of the
potential~(\ref{eqn:Vqp}) with only one frequency $\omega_1=1$, that
is a periodic potential with period 1 in $x$ and $y$, given by
\begin{equation}
V (x,y,t) =\frac{a}{2\pi}\sum_{m,n=1\atop{n^2+m^2\le N^2}}^N
\frac{1}{(n^2+m^2)^{3/2}}\sin \left[2\pi(nx + my) + \varphi_{nm} -2\pi t
\right] , \label{potential}
\end{equation}
with $N=25$.

Since the model is fluctuating in time, the eddies of Fig.~\ref{FigIIIB_1} are rapidly modified in time and where a vortex
was initially present,
an open line appears, and so on.\\
Two particular properties of the model, anisotropy and propagation
have been observed : each image of the potential field shows an
elongated structure of the eddies and superposing images obtained
at different times a slight propagation in the $y=x$ direction is
found. However, this propagation can easily be proved not to
disturb the diffusive motion of the guiding centers. The property
of propagation can be easily understood analytically. In fact,
restricting ourselves to the most simplified case of an electric
potential given only by a dominant mode ($n=m=1$) it is
immediately evident that at any given time the maxima and minima
of the sine are located on the lines $y=-x+\mathrm{constant}$. As
the amplitudes are decreasing functions of $n$ and $m$, this
structure
 is essentially preserved also in the case of many waves.
The property of anisotropy is an effect of the random phases
in producing eddies that are irregular in space.\\
\indent We notice that there are two typical time scales in the
equations of motion: the drift characteristic time $\tau_d$,
inversely proportional to the parameter $a$, and the period of
oscillation $\tau_{\omega}$ of all the waves that enter  the
potential. The competition between these two time scales
determines what kind of diffusive behaviour is
observed~\cite{pett88,pett89}. In what follows we  consider the
case of weak or intermediate chaotic dynamics (coexistence of
ordered and chaotic trajectories) which corresponds to the
quasi-linear diffusion regime~(see Sec.~\ref{diffusion of test
particles}). Whereas in the case of fully developed chaos, that
corresponds to the so-called Bohm diffusion regime, one has to
introduce a slightly more complicated approach (see remark at the
end of Sec.~\ref{sec:IIIB2}).

\subsection{Control term for a potential which varies rapidly in time}
\label{sec:IIIB2} 

We consider an electric potential of the form $V({\bf x},{\bf y},t/\epsilon)$, i.e.
such that the time scale of its variation is of order $\varepsilon$.

We consider a time-dependent Hamiltonian system described by the function
$V({\bf x},{\bf y},t/\epsilon)$, with $({\bf x},{\bf y})\in{\mathbb R}^{2L}$
canonically conjugate variables and $t$ the time.
This Hamiltonian system has $L+1/2$ degrees of freedom.

We map this Hamiltonian system with $L+1/2$ degrees of freedom to
an autonomous Hamiltonian with $L+1$ degrees of freedom by extending the phase
space from $({\bf x},{\bf y})$ to $({\bf x},{\bf y},E,\tau)$
where the new dynamical variable $\tau$ evolves as $\tau(t)=t+\tau(0)$
and $E$ is its canonically conjugate variable. The autonomous Hamiltonian
is given by
$$
H({\bf x},{\bf y},\tau,E)=E+V({\bf x},{\bf y},\tau/\epsilon).
$$
Rescaling $\tau$ by a canonical change of variable,
$$
\hat\tau=\tau/\epsilon,~~~~~\hat E=\epsilon E,
$$
one obtains
\begin{equation}
H({\bf x},{\bf y},\tau,E)=E+\epsilon V({\bf x},{\bf y},\tau),
\label{Hstream}
\end{equation}
where we have renamed $\hat E=E$ and $\hat\tau=\tau$.
In the case of rapidly time-varying potentials, Hamiltonian~(\ref{Hstream})
is in the form $H=H_0+\epsilon V$,
that is an integrable
Hamiltonian $H_0$ plus a small perturbation $\epsilon V$.

In our case $H_0=E$, i.e. independent of ${\bf x}, {\bf y}, \tau$, therefore we have
$$
\{H_0\}=\frac{\partial}{\partial{\tau}}.
$$
If we consider a potential $V({\bf x},{\bf y},\tau)$, in the periodic case we can write
$$
V({\bf x},{\bf y},\tau)=\sum_k V_k({\bf x},{\bf y})\mathe^{ik\tau},
$$
and the action of $\Gamma$, ${\mathcal R}$ and ${\mathcal N}$ on $V$ is given by
\begin{eqnarray}
&& \Gamma V=\sum_{k\neq 0} \frac{V_k({\bf x},{\bf y})}{ik}\mathe^{ik\tau}\label{gam_per},\\
&& {\mathcal R} V=V_0({\bf x},{\bf y}),\label{R_per}\\
&& {\mathcal N} V=V({\bf x},{\bf y},\tau)-V_0({\bf x},{\bf y}).
\end{eqnarray}

Otherwise, in the more general case of a non periodic potential, one can write,
under suitable hypotheses,
$$
V({\bf x},{\bf y},\tau)=\int_{-\infty}^{+\infty}\hat V({\bf x},{\bf y},k)\mathe^{ik\tau}dk,
$$
and the action of $\Gamma$, $\mathcal R$ and $\mathcal N$
operators on $V$ is given by
\begin{equation}
\label{eqnIIIB:gamma}
\Gamma V=PV\int_{-\infty}^{+\infty}\frac{\hat V({\bf x},{\bf y},k)}{ik}\mathe^{ik\tau}dk,
\label{gammaE}
\end{equation}
\begin{equation}
{\mathcal R} V=\hat V({\bf x},{\bf y},0),
\label{RE}
\end{equation}
and
\begin{equation}
{\mathcal N} V=V({\bf x},{\bf y},\tau)-\hat V({\bf x},{\bf y},0).
\end{equation}
The computation of the control term is now a straightforward
application of Eq.~(\ref{expansion_f}).\\

Following the previous section, first we
map the Hamiltonian system with $1+1/2$ degrees of freedom given
by Eq.~(\ref{eqn:Vqp}) to an autonomous Hamiltonian with two
degrees of freedom. This is obtained by extending the phase space
from $(x,y)$ to $(x,y,E,\tau)$ that is considering $E$ the
variable conjugate to the new dynamical variable $\tau$. This
autonomous Hamiltonian is
\begin{equation}
\label{eqnIIIB:H}
H(x,y,E,\tau)=E+ \sum_{m,n,k}\frac{a_k}{2\pi(n^2+m^2)^{3/2}}  \sin
[2\pi(nx+my)+\varphi_{kmn}-2\pi\omega_{k}\tau].
\end{equation}
The integrable part of the Hamiltonian from which the operators
$\Gamma$, ${\mathcal R}$ and ${\mathcal N}$ are constructed is
isochronous
$$
H_0=E.
$$
We notice that $H_0$ is resonant (since it does not depend on the
action variable $x$). From the action of $\Gamma$ and $\mathcal R$
operators computed using Eqs.~(\ref{gammaE})-(\ref{RE}), we obtain
the control term using Eq.~(\ref{expansion_f}). For instance, the
expression of $f_2$ is
\begin{eqnarray}
\lefteqn{ f_2(x,y,t)= \frac{1}{8\pi}\sum_{k,m,n\atop k',n',m'} \frac{a_k a_{k'} (n'm-nm')}{\omega_k (n^2+m^2)^{3/2}(n'^2+m'^2)^{3/2}}} \nonumber \\
&&\times\left\{ \sin [2\pi((n+n')x+(m+m')y)+\varphi_{kmn}+\varphi_{k'm'n'}-2\pi(\omega_k+\omega_{k'})t] \right. \nonumber  \\
&&\left. +\sin
[2\pi((n-n')x+(m-m')y)+\varphi_{kmn}-\varphi_{k'm'n'}-2\pi(\omega_k-\omega_{k'})t]
\right\}. \label{Vqpf2}
\end{eqnarray}

{\em Remark:} Similar calculations can be done in the case where
there is a zero frequency, e.g., $\omega_0=0$ and $\omega_k\not=
0$ for $k\not= 0$. The first term of the control term is
$$
f_2=-\frac{1}{2} \left\{\Gamma V, ({\mathcal R} +1) V\right\},
$$
where
$$
{\mathcal R}V=\sum_{m,n} \frac{a_0}{2\pi(n^2+m^2)^{3/2}} \sin
[2\pi (nx+my)+\varphi_{0mn}].
$$
\\

If we add the exact expression of the control term to $H_0+V$, the
effect on the flow is the confinement of the motion, i.e. the
fluctuations of the trajectories of the particles, around
their initial positions, are uniformly bounded for any time~\cite{vitt04}.\\
In Sec.~\ref{sec:IIIB3}, we show that truncations of the exact control term $f$, like
for instance $f_2$ or $f_2+f_3$, are able to regularize the
dynamics and to slow down the diffusion.

We notice that for the particular model (\ref{potential}) the
partial control term $f_2$ is independent of time and is given
by

\begin{figure}
\centering
\includegraphics[width=0.49\textwidth]{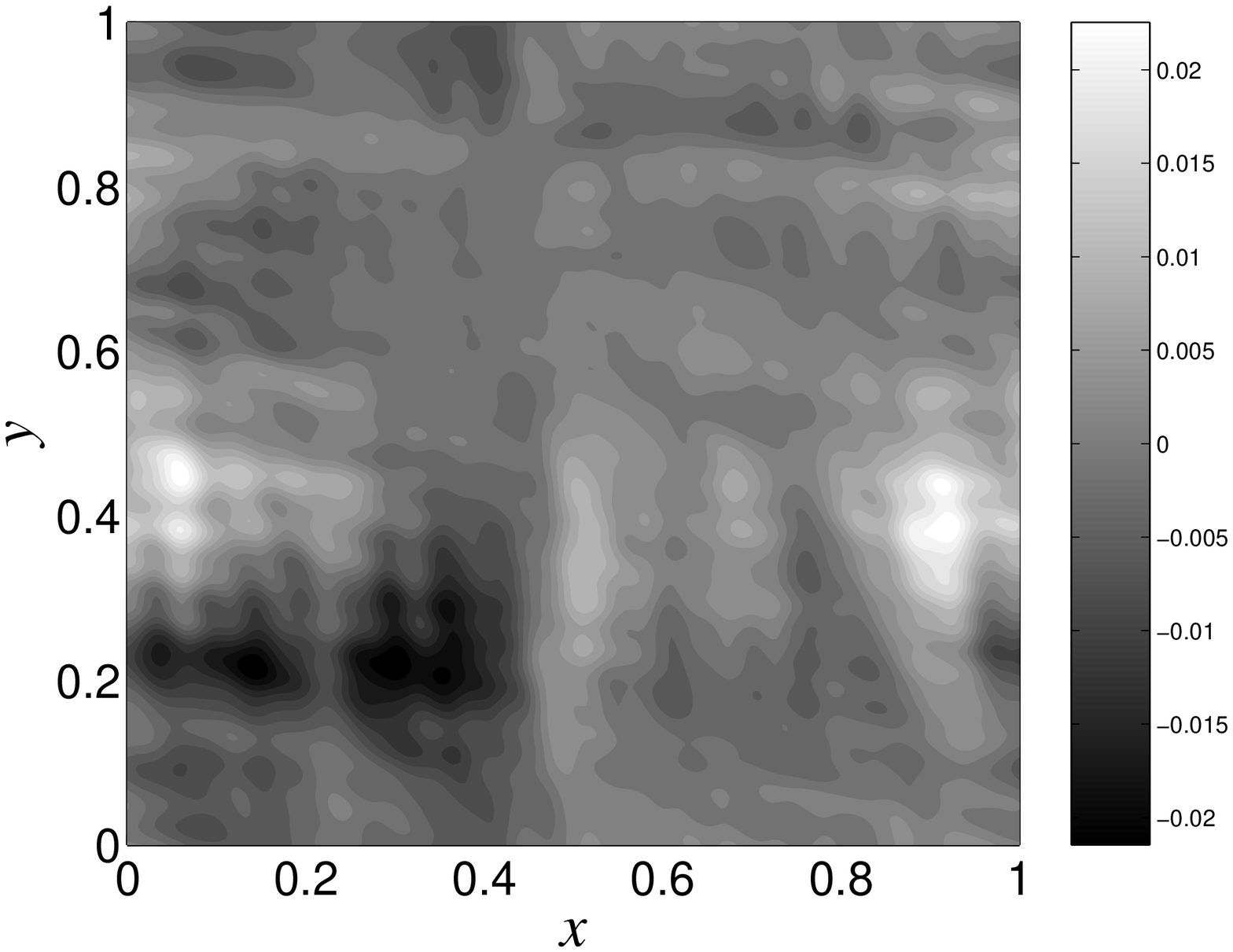}
\includegraphics[width=0.49\textwidth]{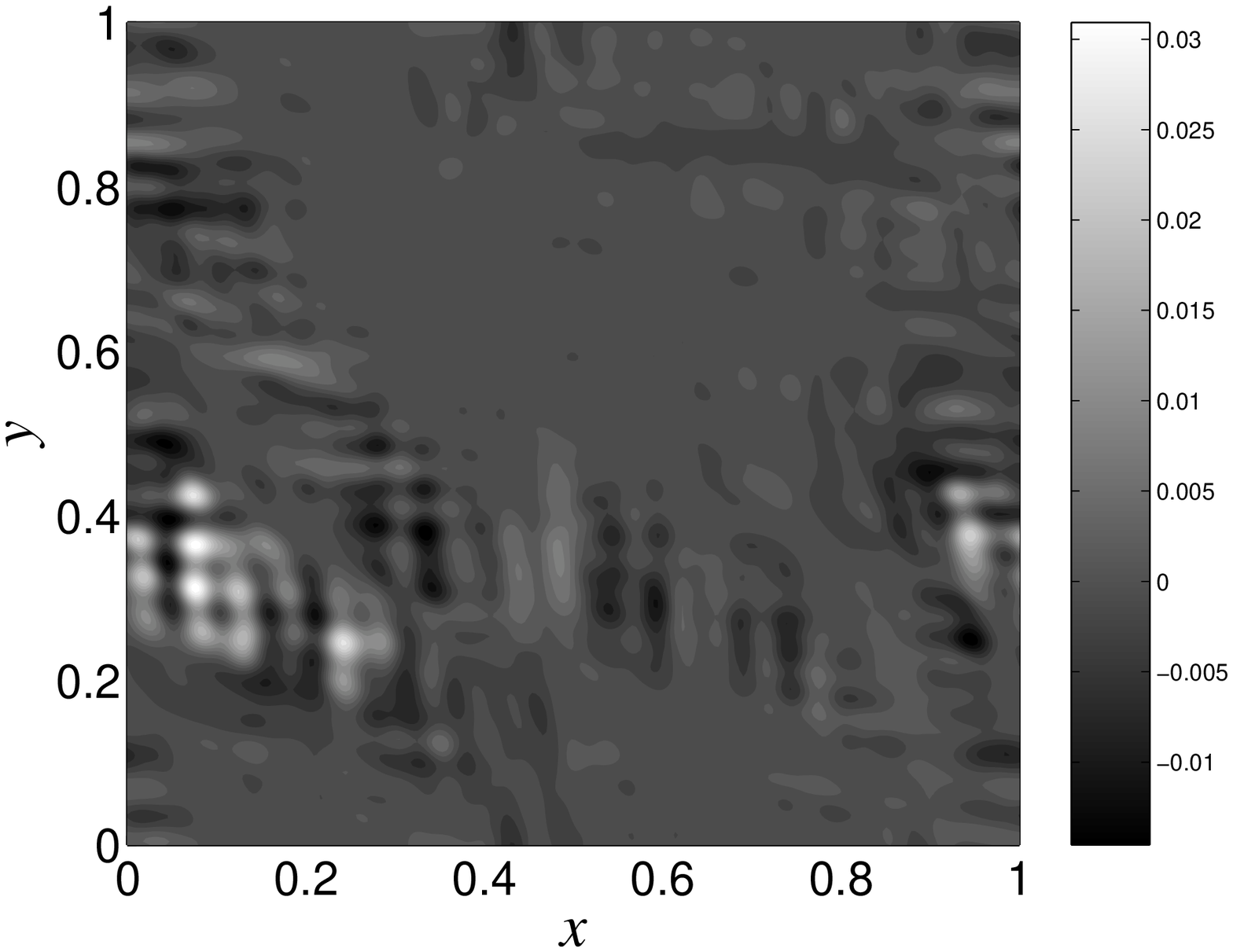}
\caption{Contour plot of $f_2$ (left panel) given by Eq.~(\ref{f_2}) and $f_3$ at $t=0$ for $a=1$
and $N=25$.} \label{FigIIIB_2}
\end{figure}
\begin{figure}
\centering
\includegraphics[width=0.49\textwidth]{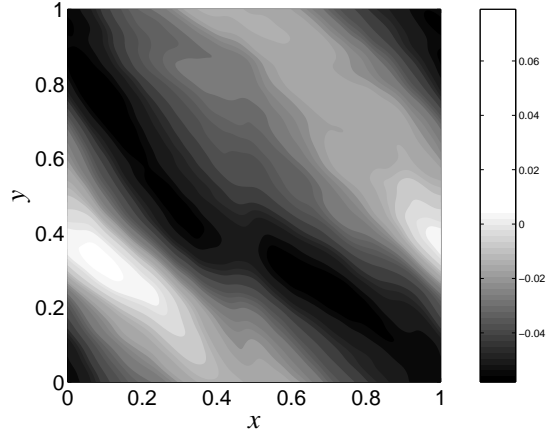}
\caption{Contour plot of $V+f_2$ for $a=0.8$
and $N=25$ at $t=1/2$.} \label{FigIIIB_2A}
\end{figure}


\begin{eqnarray}
&&f_{2}(x,y,\tau)= \frac{a^2}{8\pi}\sum_{n_1,m_1\atop{n_2,m_2}}
\frac{n_1 m_2 - n_2 m_1}{(n_1^2+m_1^2)^{3/2} (n_2^2+m_2^2)^{3/2}}  \nonumber\\
&&\times\sin \bigl[ 2\pi \bigl[ (n_1-n_2) x + (m_1-m_2) y\bigr] +
\varphi_{n_1 m_1} - \varphi_{n_2 m_2} \bigr] . \label{f_2}
\end{eqnarray}
 Figure~\ref{FigIIIB_2} depicts a contour plot of
$f_2$ given by Eq.~(\ref{f_2}). Figure~\ref{FigIIIB_2A} depicts a contour plot of $V+f_2$ for $a=0.8$ at $t=0$. We notice that the controlled potential is a small modification of the potential $V$ since this contour plot looks very similar to the one decicted in Fig.~\ref{FigIIIB_1}.

The computation of the other terms of the series
(\ref{exp_f_rv_0})
can be done recursively by using Eq.~(\ref{recursion}). Again, for $a<1$, $V+f_2+f_3$ is close to $V$, meaning that the control in this regime is based on a small modification of the potential.

\subsection{Properties of the control term} \label{sec:3c} In
this section, we state that for $a$ sufficiently small, the
exact control term exists and is regular. Then we give
estimates of the partial control terms in order to compare the
relative sizes of the different terms with respect to the
perturbation. The computations are performed for the
model~(\ref{potential})
but can be easily generalized to the more general case~(\ref{eqn:Vqp}).\\
Concerning the existence of the control term, we have the
following proposition~\cite{cira04b}~:

\begin{proposition} -- If the amplitude
$a$ of the potential is sufficiently small, there exists a control
term $f$ given by the series (\ref{exp_f_rv_0}) such that $E+V+f$
is canonically conjugate
to $E$, where $V$ is given by Eq.~(\ref{potential}).
\label{proposition3}
\end{proposition}

We have shown~\cite{cira04b} that the exact control term exists for $a\lesssim
7\times 10^{-3}$. As usual, such estimates are very conservative
with respect to realistic values of $a$. In the numerical study,
we consider values of $a$ of order 1.

Concerning the regularity of the control term of the
potential~(\ref{potential}), we notice that each term $f_s$ in the
series~(\ref{exp_f_rv_0}) is a trigonometric polynomial with an
increasing degree with $s$. The resulting control term is not
smooth but its Fourier coefficients exhibit the same power law
mode dependence as $V$:

\begin{proposition} -- All the Fourier
coefficients $f_{nmk}^{(s)} $ of the functions $f_s$ of the
series~(\ref{exp_f_rv_0}) satisfy:
\begin{equation}
\label{eqnapp3}
\vert f_{nmk}^{(s)}\vert  \leq  \frac{a^s C^s}{(n^2+m^2)^{3/2}},
\end{equation}
for $(n,m)\not= (0,0)$. Consequently, for $a$ sufficiently small,
the Fourier coefficients of the control term $f$ given by
Eq.~(\ref{exp_f_rv_0}) satisfy:
$$
\vert f_{nmk}\vert  \leq  \frac{C_\infty}{(n^2+m^2)^{3/2}},
$$
for $(n,m)\not= (0,0)$ and for some constant $C_\infty >0$.
\label{proposition4}
\end{proposition}

\subsection{Magnitude of the control term}
A measure of the relative sizes of the control terms is defined by
the electric energy density associated with each electric field
$V$, $f_2$ and $f_3$. From the potential we get the electric field
and hence the motion of the particles. We define an average energy
density ${\mathcal E}$ as
$$
{\mathcal E}=\frac{1}{8\pi}\langle~ \Vert{\bf E}\Vert^2~\rangle,
$$
where ${\bf E}(x,y,t)=-{\bf\nabla}V$. In terms of the particles,
it corresponds to the mean value of the kinetic energy $\langle
\dot{x}^2+\dot{y}^2\rangle$ (up to a multiplicative constant). For
$V(x,y,t)$ given by Eq.~(\ref{potential}),
\begin{equation}
{\mathcal E}=\frac{a^2}{8\pi}\sum_{n,m=1\atop{n^2+m^2\le N^2}}^N
\frac{1}{(n^2+m^2)^2}. \label{enepotenziale}
\end{equation}
We define the contribution of $f_2$ and $f_3$ to the energy
density by
\begin{equation}
e_2=\frac{1}{8\pi} \langle \Vert {\bf \nabla}f_2 \Vert^2\rangle.
\label{e2}
\end{equation}
For $N=25$, these contributions satisfy:
$$
\frac{e_2}{{\mathcal E}}\approx 0.1\times a^2.
$$
It means that the control terms $f_2$ can be considered
as small perturbative terms with respect to $V$ when $a<1$.\\

{\em Remark on the number of modes in $V$}: In this section, all
the computations have been performed with a fixed number of modes
$N=25$ in the potential $V$ given by Eq.~(\ref{potential}).
The question we address in this remark is how the results are
modified as we increase $N$. First we notice that the potential
and its electric energy density are bounded with $N$ since
\begin{eqnarray*}
&& \vert V(x,y,t)\vert \leq a \sum_{n,m=1}^\infty \frac{1}{(n^2+m^2)^{3/2}} < \infty ,\\
&& {\mathcal E}\leq \frac{a^2}{8\pi}\sum_{n,m=1}^\infty
\frac{1}{(n^2+m^2)^2}< \infty .
\end{eqnarray*}
Concerning the partial control term $f_2$, we see that it is in
general unbounded with $N$. From its explicit form it grows like $N\log N$. Less is known on
the control term since it is given by a series whose terms are
defined by recursion. However, from Proposition~\ref{proposition4}, we can show that the value $a$ of existence
of the control term decreases like $1/(2N\log N)$. This divergence
of the control term comes from the fact that the Fourier
coefficients of the potential $V$ are weakly decreasing with the
amplitude
of the wavenumber.\\
Therefore, the exact control term might not exist if we increase
$N$ by keeping $a$ constant. However for practical purposes the Fourier series of the control term
can be truncated to its first terms (the Fourier modes with
highest amplitudes). Furthermore in the example we consider as
well as for any realistic situation the value of $N$ is bounded by
the resolution of the potential. In the case of electrostatic
turbulence in plasmas $k\rho_i\sim 1$  determines an upper bound
for $k$, where $k$ is the transverse wave vector related to the
indices $n,m$ and $\rho_i$ the ion Larmor radius. The physics
corresponds to the averaging effect introduced
by the Larmor rotation.\\

\subsection{Efficiency and robustness}
\label{sec:IIIB3} With the aid of numerical simulations (see
Refs.~\cite{pett88,pett89} for more details on the numerics), we
check the effectiveness of the control term by comparing the
dynamics of particles  obtained from the uncontrolled Hamiltonian
and from the same Hamiltonian with the control term $f_2$ and with
a more refined control term $f_2+f_3$. 

\subsubsection{Diffusion of test particles}
\label{diffusion of test particles} 
\begin{figure}
\includegraphics[width=0.49\textwidth]{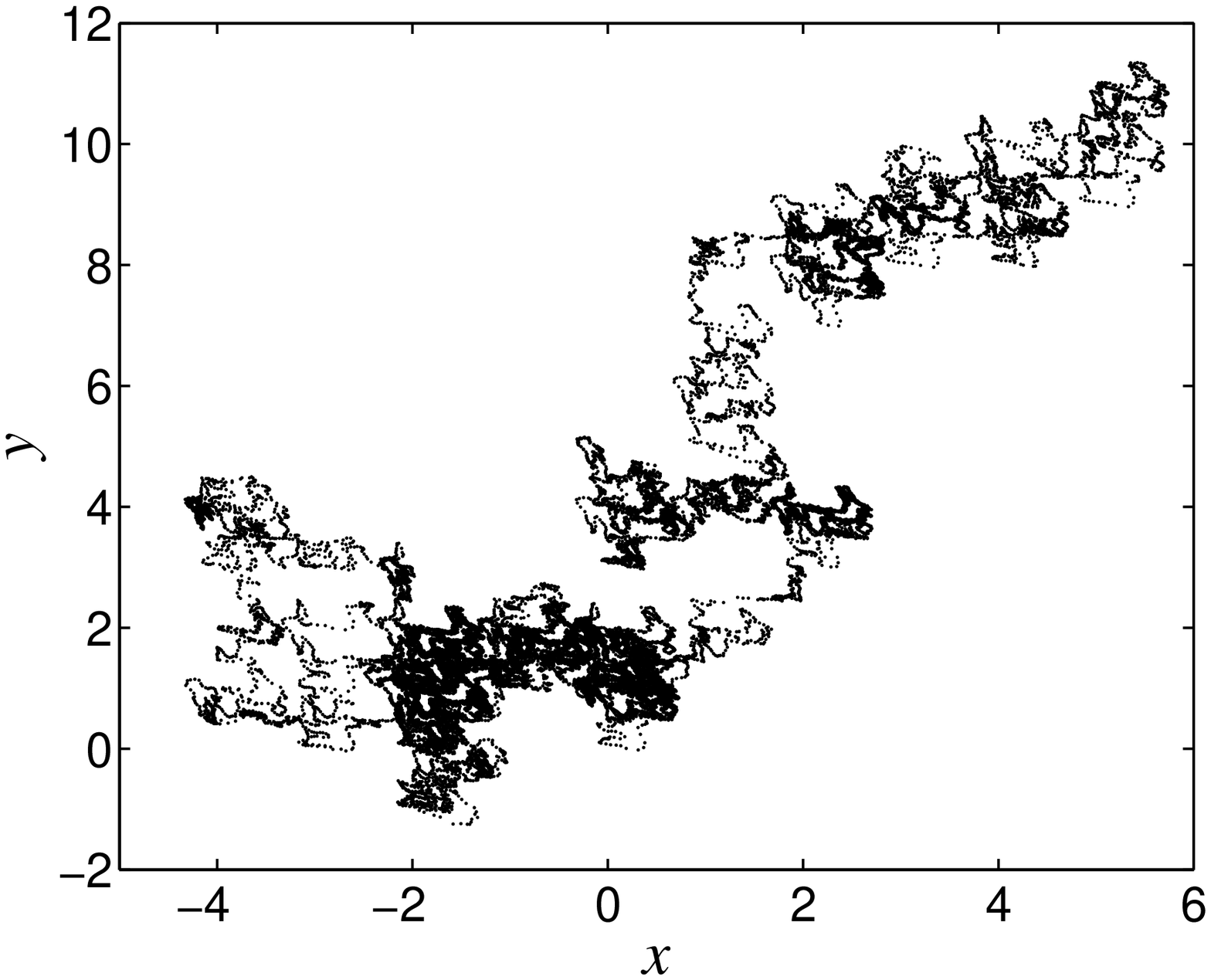}
\includegraphics[width=0.49\textwidth]{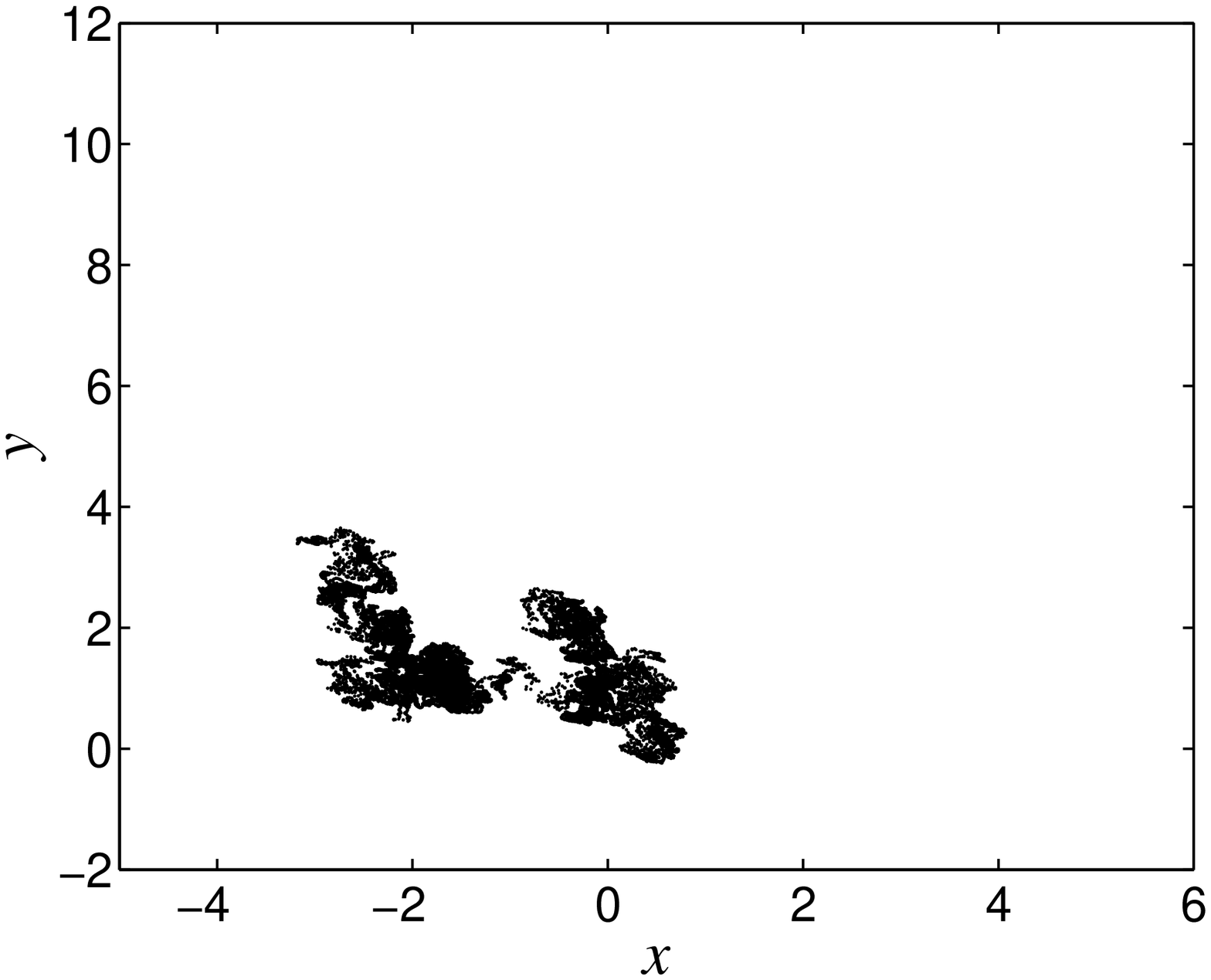}
\caption{\label{FigIIIB_4}Poincar\'e sections of a
trajectory obtained using a generic initial condition for
Hamiltonian (\ref{potential})  with $a=0.8$, without control term
(upper left panel) and with control term (\ref{f_2}) (upper
right panel).} 
\end{figure}

The effect of the control
terms can first be seen from a few randomly chosen trajectories.
We have plotted Poincar\'e sections (stroboscopic plots of the
trajectories of $V$) on Fig.~\ref{FigIIIB_4} of two trajectories
issued from generic initial conditions computed without and with
the control term $f_2$ respectively. Similar pictures are obtained
for many other randomly chosen initial conditions. The stabilizing
effect of the control term (\ref{f_2}) is illustrated by such
trajectories. The motion remains diffusive but the extension of
the phase space explored by the trajectory is reduced.

\begin{figure}
\centering
\includegraphics[width=0.49\textwidth]{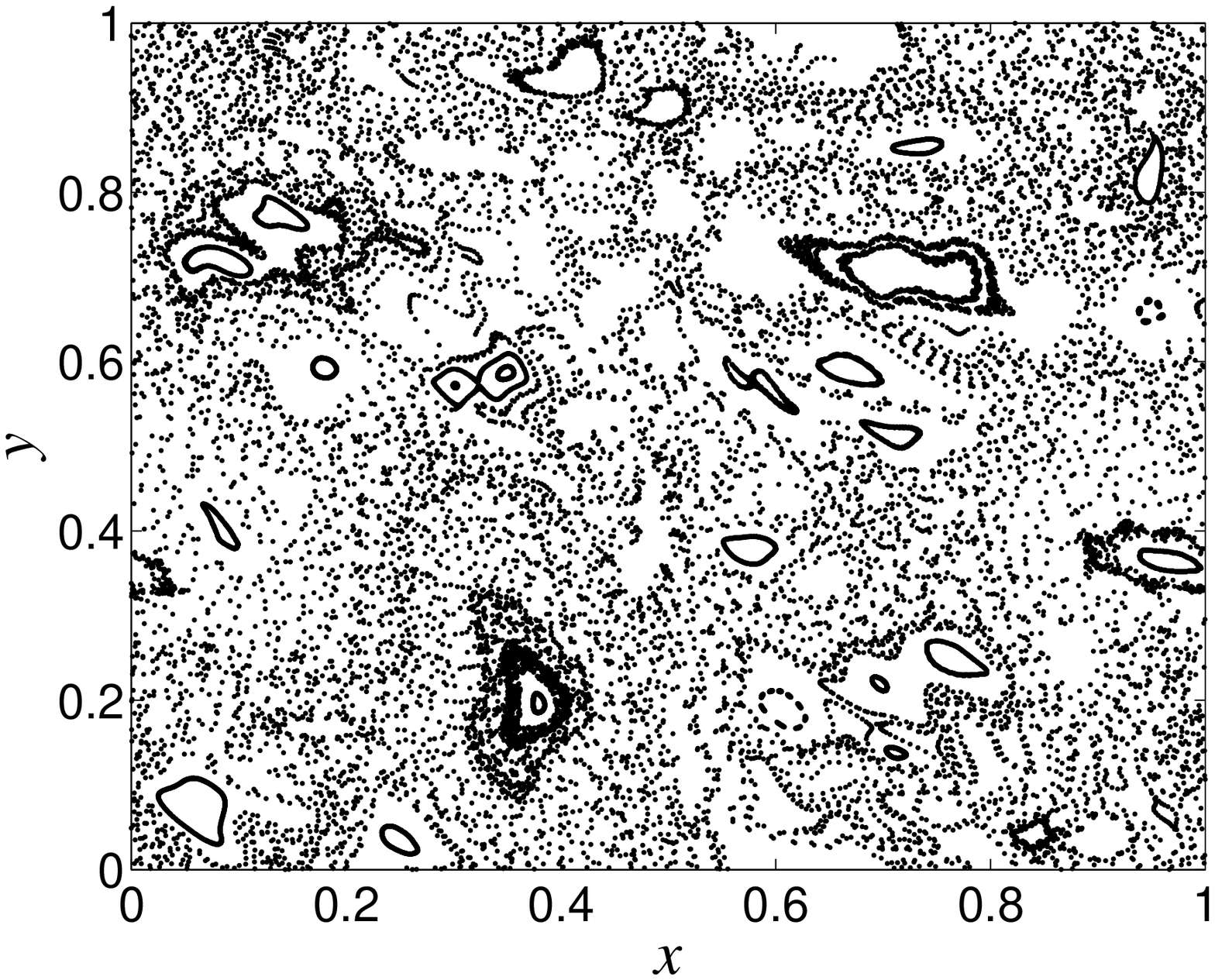}
\includegraphics[width=0.49\textwidth]{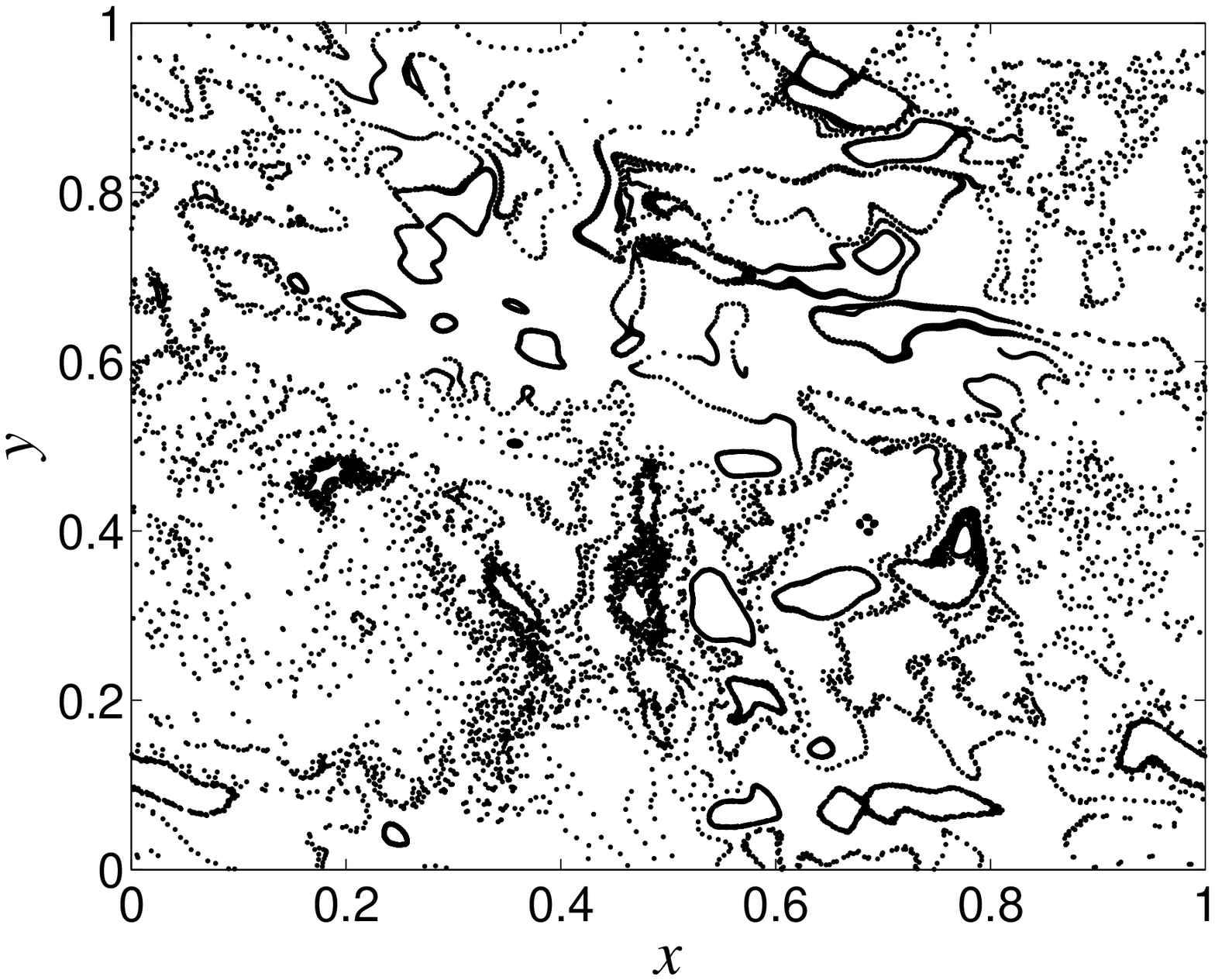}
\includegraphics[width=0.49\textwidth]{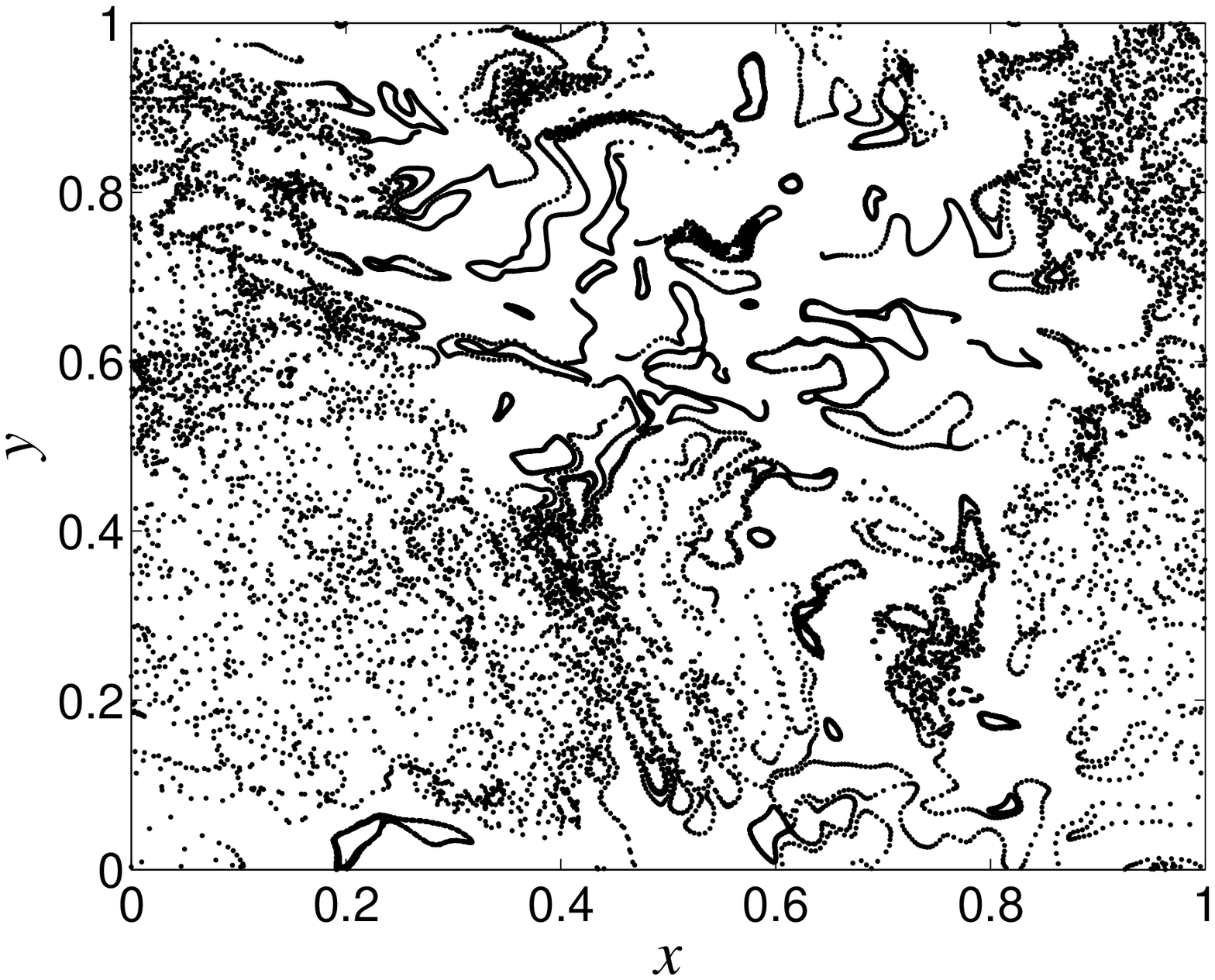}
\caption{Poincar\'e sections on the $[0,1]^2$ torus for $V$ (upper left panel), $V+f_2$ (upper right panel) and $V+f_2+f_3$ (lower panel) with $a=0.6$} \label{FigIIIB_4PS}
\end{figure}

The dynamics is more clearly seen on a Poincar\'e section on the $[0,1]^2$ torus (i.e.\ by taking $x$ and $y$ modulo 1). Such Poincar\'e sections are depicted in Fig.~\ref{FigIIIB_4PS} for $V$ (upper left panel), $V+f_2$ (upper right panel) and $V+f_2+f_3$ (lower panel). These figures shows that the Poincar\'e sections are composed of two types of trajectories~: ones which are trapped around resonant islands, and diffusive (chaotic) ones which lead to global transport properties. We notice that the number of resonant islands has been drastically increased by the control term. The mechanism of reduction of chaos seems to be that a significant number of periodic orbits has been stabilized by the addition of the control term.

\begin{figure}
\includegraphics[width=0.49\textwidth]{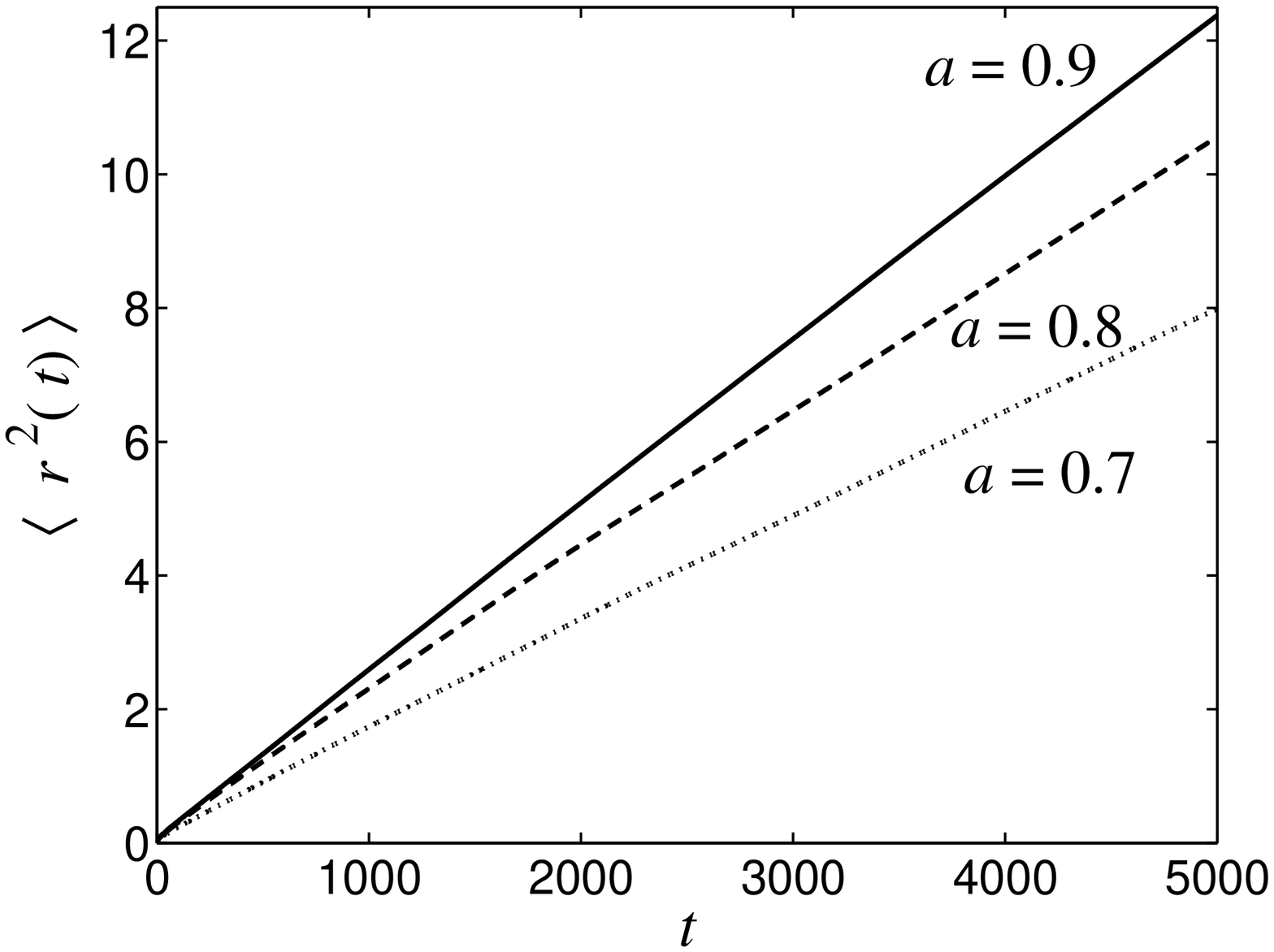}
\includegraphics[width=0.49\textwidth]{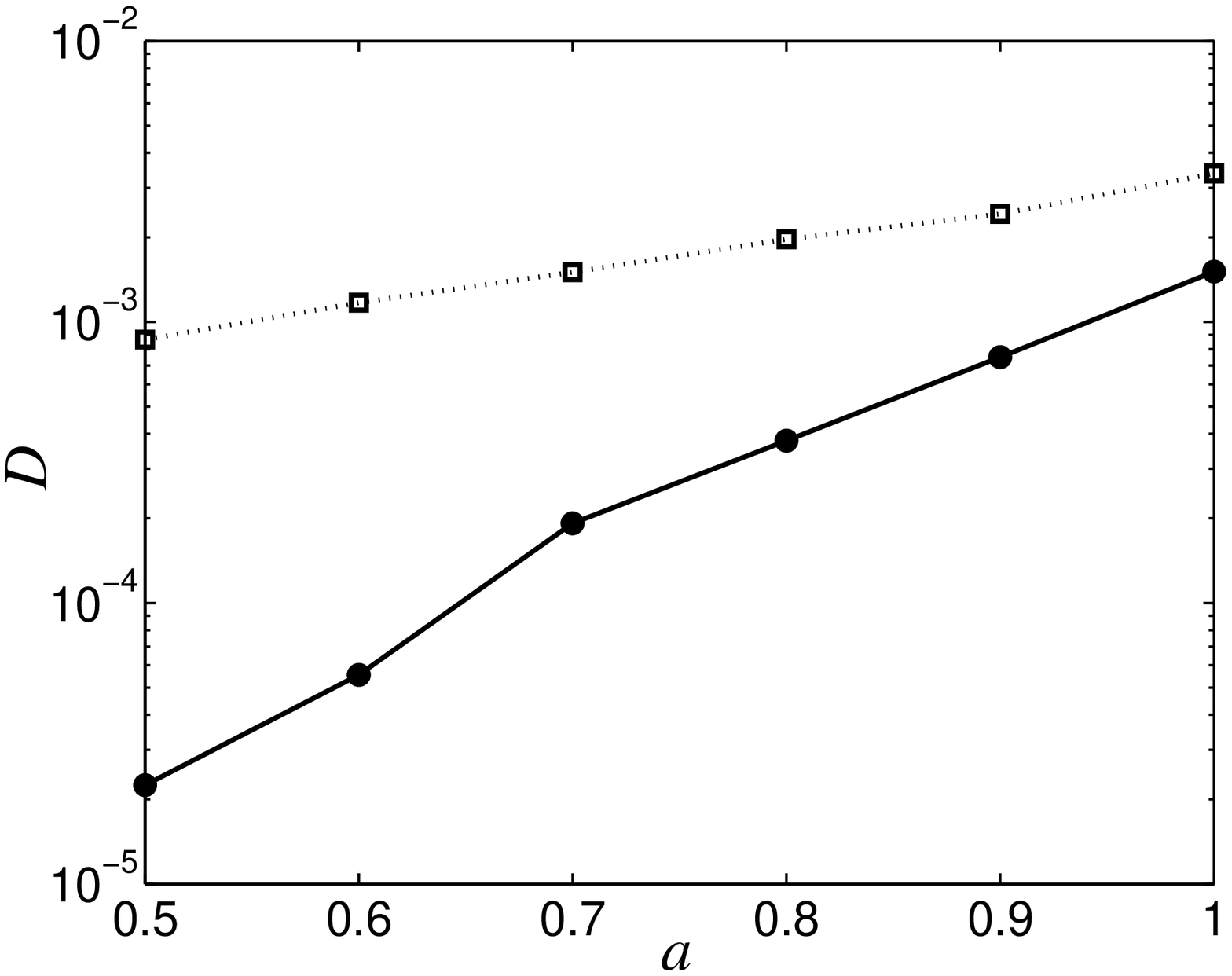}
\caption{On the left panel, mean square displacement $\langle r^2
(t)\rangle $ versus time $t$ obtained for  Hamiltonian
(\ref{potential}) with three different values of $a=0.7$, $a=0.8$,
$a=0.9$. In the right panel, diffusion coefficient $D$ versus $a$
obtained for Hamiltonian (\ref{potential}) (open squares) and
Hamiltonian (\ref{potential}) plus control term (\ref{f_2}) (full
circles).} \label{FigIIIB_5}
\end{figure}

A clear evidence is found for a relevant reduction of the
diffusion in presence of the control term (\ref{Vqpf2}).
In order to study the diffusion properties of the system, we have
considered a set of $\mathcal M$ particles (of order $1000$)
uniformly distributed at random in the domain $0\leq x,y\leq 1$ at
$t=0$. We have computed the mean square displacement $\langle r^2
(t) \rangle$ as a function of time
$$
\langle r^2 (t) \rangle = \frac{1}{\mathcal M}
\sum_{i=1}^{\mathcal M} {\Vert{\bf x}_i(t) - {\bf x}_i(0)\Vert}^2
$$
where ${\bf x}_i(t)=(x_i(t),y_i(t))$ is the position of the
$i$-th particle at time $t$ obtained by integrating Hamilton's
equations with initial conditions ${\bf x}_i(0)$. Figure~\ref{FigIIIB_5} (left panels) shows $\langle r^2 (t) \rangle$ for three
different values of $a$. For the range of parameters we consider,
the behavior of $\langle r^2 (t)\rangle$ is always found to be
linear in time for $t$ large enough. The corresponding diffusion
coefficient is defined as
\[
D= \lim_{t \rightarrow\infty}{{\langle r^2 (t) \rangle} \over t}~.
\]

Figure~\ref{FigIIIB_5} (right panel) shows the values of $D$ as
a function of $a$ with and without control term. It clearly shows
a significant decrease of the diffusion coefficient when the
control term is added. As expected, the action of the control term
gets weaker as $a$ is increased towards the strongly chaotic
phase. We notice that the diffusion coefficient is plotted on a log-scale. For $a=0.7$, the control reduces the diffusion coefficient by a factor approximately equal to 10.

\subsubsection{Robustness of the control} 
\label{sec:IIIB5} In the
previous sections, we have seen that a truncation of the series
defining the control term by considering the first $f_2$ or the two
first terms $f_2+f_3$ in the perturbation series in the small parameter
$a$,
gives a very efficient control on the chaotic dynamics of the system. This reflects the robustness of the method.\\
In this section we show that  it is possible to use other types of approximations of the control term
and still
get an efficient control of the dynamics.

\begin{figure}
\centering
\includegraphics[width=0.49\textwidth]{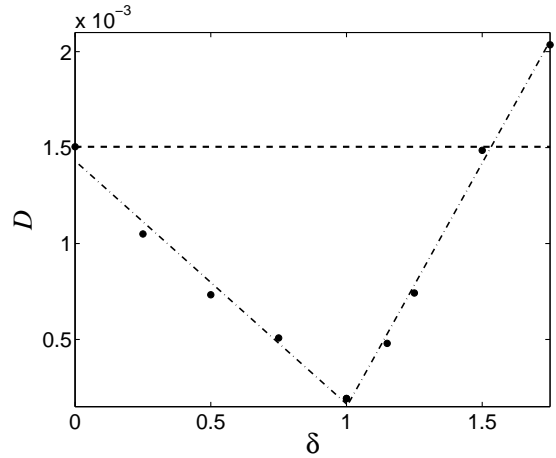}
\caption{Diffusion coefficient $D$ versus the magnitude of the
control term $\delta\, f_2$ where $f_2$ is given by Eq.~(\ref{f_2}) for $a = 0.7$. The horizontal dashed line
corresponds to the value of $D$ without control term ($\delta=0$). The
dash-dotted line is a piecewise linear interpolation.}
\label{FigIIIB_18}
\end{figure}

We check the robustness of the control by increasing or reducing
the amplitude of the control term~\cite{cira04b}. We replace $f_2$ by
$\delta\cdot f_2$ and we vary the parameter $\delta$ away from its
reference value $\delta =1$. Figure~\ref{FigIIIB_18} shows that both
the increase and the reduction of the magnitude of the control
term (which is proportional to $\delta\cdot a^2$) result in a loss
of efficiency in reducing the diffusion coefficient. The fact that
a larger perturbation term -- with respect to the computed one --
does not work better, also means that the control is not a ``brute force'' effect.\\ The interesting result
is that one can significantly reduce the amplitude of the control
($\delta <1$) and still get a reduction of the chaotic diffusion.
We notice that the average energy density $e_2(\delta)$ associated
with a control term $\delta \cdot f_2$ is equal to
$e_2(\delta)=\delta^2 e_2$, where $e_2$ is given by
Eq.~(\ref{e2}).
Therefore, for $\delta=0.5$ where the energy necessary for the
control is one fourth of the optimal control, the diffusion
coefficient is significantly smaller than in the uncontrolled
case $\delta=0$~(by nearly a factor $3$).


\end{document}